\documentstyle[amssymb,preprint,aps,prb]{revtex}

\begin{document}
\title{FAST INCOMPLETE DECOHERENCE OF NUCLEAR SPINS IN QUANTUM HALL FERROMAGNET}
\author{T.Maniv$^{1,2}$ , Yu. A. Bychkov$^{1,3}$, I.D. Vagner$^{1}$, and P. Wyder$%
^{1}$}
\address{$^{1}$Grenoble High Magnetic Field Laboratory,Max-Planck-\\
Institute fur Festkorperforschung and CNRS, Grenoble\\
France.\\
$^{2}$Chemistry Department, Technion-Israel Institute of\\
Technology, Haifa 32000, Israel\\
$^{3}$L.D.Landau Institute for Theoretical Physics, Kosygina 2, Moscow,\\
Russia}
\date{\today{}}
\maketitle

\begin{abstract}
A scenario of quantum computing process based on the manipulation of a large
number of nuclear spins in Quantum Hall (QH) ferromagnet is presented. It is
found that vacuum quantum fluctuations in the QH ferromagnetic ground state
at filling factor $\nu =1$ , associated with the virtual excitations of spin
waves, lead to fast incomplete decoherence of the nuclear spins. A
fundamental upper bound on the length of the computer memory is set by this
fluctuation effect.

PACS numbers:
\end{abstract}

A growing number of models for quantum information processing (or Quantum
Computing-QC) has been recently proposed \cite{Steane98}, some of which were
successfully tested experimentally in devices consisting of a few qbits.The
scaling up of these toy devices to the desired large number of quantum gates
seems at present a formidable challenge. Of special interest are the models
based on the manipulation of nuclear spins in semiconducting
heterostructures \cite{PVK98},\cite{Kane98}. A scenario for the realization
of QC over a large number of qbits in a model system similar to that
proposed in Ref.[\cite{PVK98}] may be achieved when nuclear spins in
heterojunctions are manipulated via the hyperfine interaction with the
electron spins under the conditions of the odd integer Quantum Hall (QH)
effect.

The main idea behind this scenario is motivated by the experimental
observation in optically pumped NMR measurements on GaAS multiple quantum
well structure at low temperatures of a dramatic enhancement of the very
small nuclear spin lattice relaxation rate which charaterizes the QH
ferromagnetic state at Landau level filling factor $\nu =1$ ( i.e.
corresponding to $T_{1}>250$ $s$ ), and of a sharp decrease of the Knight
shift , as the filling factor is shifted slightly away from $\nu =1$ [\cite
{Tycko95}]. The prevailing interpretation of these closely related effects,
associates them with the creation of skyrmions (or antiskyrmions) in the
electron spin distribution as the 2D electron system moves away from the QH
ferromagnetic state at $\nu =1$ [\cite{Girvin99}].

It should be stressed, however, that the nuclear spin dephasing time $T_{2}$
in quantum well structures based on GaAS/AlGaAs, is expected to be of the
order of milliseconds or shorter, namely much smaller than the shortest
value of $T_{1}$ found in this experiments. This drawback is due to the fact
that all elemental components of such structures (i.e. Ga$^{69}$ , Ga$^{71}$
, As$^{75}$ , all with $I=3/2$ , and Al$^{27}$ with $I=5/2$ ) have non-zero
nuclear spins, which as a result experience large direct dipolar
interactions. An alternative quantum well structure based on semiconducting
host material consisting predominantly of zero nuclear spin isotopes and
small amount of atoms with nonzero nuclear spins (e.g. like Si/Si$_{1-x}$Ge$%
_{x}$ heterojunctions \cite{shlimak83}), may be fabricated in the future to
create a system of virtually {\it noninteracting} nuclear spins in a QH
ferromagnet.

In such a system the nuclear dephasing time $T_{2}$ is governed only by the
hyperfine interactions with the electron spins. Random impurities, for
example, can influence this nuclear spin dephasing only indirectly through
the scattering of electrons by the impurity potential, leading ,e.g. in
typical GaAs quantum well structure to nuclear spin dephasing times of the
order of seconds ( see below ) \cite{MPV00} .

One therefore expect that after saturating the nuclear spins and tuning the
filling factor at $\nu =1$ a nearly pure nuclear spin state can be frozen
for a relatively long time due to the very long nuclear spin relaxation
times $T_{1}$ and $T_{2}$. The nuclear spin system can be now manipulated by
varying the filling factor away from the initial value and then back to $\nu
=1$ to freeze in a new configuration. During this variation the 2D electron
system is in a condensed state of a large number of spin waves (or spin
excitons) \cite{BMV96}, which are strongly correlated over a large spatial
region. It is thus expected that the proposed manipulation of the nuclear
spin system can be performed in a phase coherent fashion over a spatial
region with size of the order of the skyrmion radius.

In attempting to evaluate the feasibility of this hypothetical scenario
several problems of various degrees of complexity arise. Perhaps, the most
fundamental one concerns the mechanism in which a single nuclear spin (or
qbit) looses phase coherence by the 'manipulating agent' itself , i.e. the
2D electron system in the heterostructure. In particular, it is not at all
obvious that even under the ideal conditions of the QH ferromagnetic state
at $\nu =1$ , where the nuclear spin relaxation times, $T_{1}$ and $T_{2}$,
are extremely long, an ensemble of large number of nuclear spins can
preserve phase coherence for a sufficiently long time.

Of special interest here is the effect of vacuum quantum fluctuations in the
QH ferromagnetic state on the decoherence of nuclear spins. As will be shown
below, the virtual excitations of spin waves (or spin excitons \cite{BIE81}, 
\cite{KH84}), which have a large energy gap (on the scale of the nuclear
Zeeman energy) above the ferromagnetic ground state energy, lead to fast
incomplete decoherence in the nuclear spin system.

To study this effect we exploit an ideal model system consisting of
independent nuclear spins interacting with a 2D electron gas through the
Fermi contact hyperfine interaction. An external stationary magnetic field
is applied perpendicular to the 2D layer with strength corresponding to
filling factor $\nu =1$. The temperature is assumed to be smaller than any
electronic energy scale in the problem , and the influence of electron
scattering by impurities is neglected.

The Hamiltonian of this model system is \ \ \ $\widehat{H}=\widehat{H}_{0}+%
\widehat{H}_{en}$ \ , \ \ where 
\begin{equation}
\widehat{H}_{0}=-\gamma _{n}\sum_{j}\widehat{{\bf I}}_{j}\cdot {\bf B}%
_{0}-\gamma _{e}\int d^{2}r\widehat{{\bf S}}\left( {\bf r}\right) \cdot {\bf %
B}_{0}+\widehat{H}_{ee}  \label{Ho}
\end{equation}
\begin{equation}
\widehat{H}_{en}=A\sum_{j}\widehat{{\bf S}}\left( {\bf r}_{j}\right) \cdot 
\widehat{{\bf I}}_{j}  \label{Hint}
\end{equation}

Here $\widehat{{\bf I}}_{j}$ is the nuclear spin operator located at ${\bf r}%
_{j}$ , $\widehat{{\bf S}}\left( {\bf r}\right) $ is the electronic spin
density operator , ${\bf B}_{0}$ is the external magnetic field which is
assumed to be oriented perpendicular to the 2D EG ( ${\bf B}_{0}=B_{0}{\bf z}
$ ) , $\widehat{H}_{ee}$ is the electron electron interaction, $\gamma _{n}$
and $\gamma _{e}$ the nuclear and electronic gyromagnetic ratios
respectively,

\[
A=C\left| u\left( 0\right) \right| ^{2}/ll_{B}^{2} 
\]
with $C=\frac{8\pi }{3}\gamma _{n}\gamma _{e}$ , and $u\left( 0\right) $ the
electron wavefunction at a nucleus. The difference between the Fermi contact
hyperfine interaction parameter $A$ in a quantum well at high magnetic field
and the corresponding zero field bulk coupling constant is reflected in the
appearance of the two length parameters; $l$ -the width of the quantum well
, and $l_{B}=\sqrt{c\hbar /eB_{0}}$ - the magnetic length. \ The zero field
bulk value, 
\[
A_{Bulk}=C\left| u\left( 0\right) \right| ^{2}/\Omega 
\]
where $\Omega $ is the volume of a unit cell , is usually much larger than $%
A $ given above as long as $ll_{B}^{2}\gg \Omega $ .

\smallskip The manipulation of the nuclear spins is carried out through spin
flip-flop processes, associated with the 'transverse' part of the
interaction Hamiltonian $\widehat{H}_{en}$ ( Eq.(\ref{Hint}) ) , i.e. 
\[
A\sum_{j}\left[ \widehat{I}_{j,+}\widehat{S}_{-}\left( {\bf r}_{j}\right) +%
\widehat{I}_{j,-}\widehat{S}_{+}\left( {\bf r}_{j}\right) \right] 
\]
where $\widehat{I}_{j,+}=\widehat{I}_{j,x}+i\widehat{I}_{j,y},\widehat{I}%
_{j,-}=\widehat{I}_{j,x}-i\widehat{I}_{j,y}$ \ are the transverse components
of the nuclear spin operators, and $\widehat{S}_{+}\left( {\bf r}\right) =%
\widehat{\psi }_{\downarrow }^{\dagger }\left( {\bf r}\right) \widehat{\psi }%
_{\uparrow }\left( {\bf r}\right) $ , $\widehat{S}_{-}\left( {\bf r}\right) =%
\widehat{\psi }_{\uparrow }^{\dagger }\left( {\bf r}\right) \widehat{\psi }%
_{\downarrow }\left( {\bf r}\right) $ are the corresponding components of
the electron spin density operators. Here $\widehat{\psi }_{\sigma }\left( 
{\bf r}\right) ,\widehat{\psi }_{\sigma }^{\dagger }\left( {\bf r}\right) $
are the electron field operators with spin projections $\sigma =\uparrow
,\downarrow $ .

The 'longitudinal' part of $\widehat{H}_{en}$, $A\sum_{j}\widehat{I}_{j,z}%
\widehat{S}_{z}\left( {\bf r}_{j}\right) $ , which commutes with the
Hamiltonian $\widehat{H}_{0}$ , and so leaves the nuclear spin projections
along ${\bf B}_{0}$ unchanged, can still erodes quantum coherence in the
nuclear spin system \cite{Palma96}.

To simplify the analysis we assume that the nuclei under study have spin $1/2
$ . In this case the transverse components $\widehat{I}_{j,+}$ , $\widehat{I}%
_{j,-}$ , are up to a proportionality constant, just the off diagonal
elements (or coherences) of the density matrix of a single nuclear spin
(qbit) \cite{Cohentan, Slichter90}. The decay of these elements with time,
which determines the rate of decoherence of a single qbit, can be thus found
from \bigskip the equations of motion for the operators $\widehat{I}_{j,\pm
}\left( t\right) $ in the Heisenberg representation $\widehat{I}_{j,\pm
}\left( t\right) =e^{i\widehat{H}t/\hbar }\widehat{I}_{j,\pm }e^{-i\widehat{H%
}t/\hbar }$ .

Let us, for the sake of simplicity, consider a single nuclear spin and
evaluate its rate of decoherence due to the coupling with a 'bath' of spin
excitons. Dropping the site index $j$ , the corresponding equations for the
coherences $I_{+}$ , $I_{-}$ \ , can be written in the form : 
\begin{equation}
\frac{\partial }{\partial t}\widetilde{I}_{+}\left( t\right) =-i\alpha 
\widetilde{S}_{+}\left( t\right) \widehat{I}_{z}\left( t\right) \;,\;\frac{%
\partial }{\partial t}\widetilde{I}_{-}\left( t\right) =i\alpha \widetilde{S}%
_{-}\left( t\right) \widehat{I}_{z}\left( t\right)  \label{d/dtI_tr}
\end{equation}
with the supplementary equation for $\widehat{I}_{z}\left( t\right) $%
\begin{equation}
\frac{\partial }{\partial t}\widehat{I}_{z}\left( t\right) =\frac{i}{2}%
\alpha \left[ \widetilde{S}_{+}\left( t\right) \widetilde{I}_{-}\left(
t\right) -\widetilde{S}_{-}\left( t\right) \widetilde{I}_{+}\left( t\right) %
\right]  \label{d/dtI_z}
\end{equation}

Here $\alpha \equiv A/\hbar $ , and the symbols $\widetilde{I}_{\pm }\left(
t\right) $ , $\widetilde{S}_{\pm }\left( t\right) $ stand for the
corresponding spin operators\ in the rotating reference of frame with
angular velocity $\omega =\gamma _{n}B_{0}-AS_{z}$ [\cite{Slichter90}] ,
i.e. : 
\[
\widetilde{I}_{\pm }\left( t\right) \equiv e^{\pm i\omega t}\widehat{I}_{\pm
}\left( t\right) \;;\;\widetilde{S}_{\pm }\left( t\right) \equiv e^{\pm
i\omega t}\widehat{S}_{\pm }\left( t\right) 
\]

Assuming that initially, at time $t=0$ , the electronic system is in its
ground (QH ferromagnetic) state $|0\rangle $ , and neglecting the effect of
the nuclear spins on the electronic (bath) states, the average of Eq.(\ref
{d/dtI_tr}) over the 'bath' states reduces to the expectation value in the
ground electronic state $|0\rangle $. Thus, by integrating Eq.(\ref{d/dtI_z}%
) over $t$ , substituting into Eq.(\ref{d/dtI_tr}), and then averaging over
the 'bath' states, one finds to lowest order in the hyperfine interaction
parameter $\alpha $ : 
\begin{equation}
\frac{\partial }{\partial t}\widetilde{I}_{+}\left( t\right) =-\frac{1}{2}%
\alpha ^{2}\int_{0}^{t}d\tau e^{i\omega _{n}\tau }\chi _{+-}\left( \tau
\right) \widetilde{I}_{+}\left( t-\tau \right)  \label{d/dtI_+}
\end{equation}
where $\chi _{+-}\left( \tau \right) \equiv \left\langle 0\right| \widehat{S}%
_{+}\left( t\right) \widehat{S}_{-}\left( t-\tau \right) \left|
0\right\rangle $ .

Note that since $\chi _{+-}\left( \tau \right) $ varies on the
characteristic {\it electronic} time scale $1/\gamma _{e}B_{0}$ , which is
much shorter than the {\it nuclear} time scale $\omega ^{-1}$ , it is
allowed to use the expansion $\widetilde{I}_{+}\left( t-\tau \right) =%
\widetilde{I}_{+}\left( t\right) -\tau \frac{d}{dt}\widetilde{I}_{+}\left(
t\right) +...$ under the integral in Eq.(\ref{d/dtI_+}) to get: 
\begin{equation}
\frac{\partial }{\partial t}\widetilde{I}_{+}\left( t\right) =-\frac{1}{2}%
\alpha ^{2}\int_{0}^{t}d\tau e^{i\omega \tau }\chi _{+-}\left( \tau \right) 
\widetilde{I}_{+}\left( t\right) +O(A^{4})  \label{d/dtI_+appr}
\end{equation}

The 'bath' excited states are spin excitons with the well known dispersion
relation $E_{ex}\left( k\right) $ [\cite{BIE81}], which in the long
wavelength limit reduces to $E_{ex}\left( k\right) \approx \varepsilon _{sp}+%
\frac{1}{4}\varepsilon _{c}l_{B}^{2}k^{2}$ , with $\varepsilon _{c}=\frac{%
e^{2}}{\kappa l_{B}}\sqrt{\frac{\pi }{2}}$ the characteristic Coulomb
energy. Here $\kappa $ is the dielectric constant in the 2D electron gas
region. The energy gap is the 'bare' Zeeman energy $\varepsilon _{sp}=$ $%
\hbar \gamma _{e}B_{0}$ .

A simple calculation yields for the correlation function $\chi _{+-}\left(
\tau \right) =\sum_{{\bf k}}e^{-\frac{1}{2}\widetilde{k}^{2}}e^{-iE_{ex}%
\left( k\right) \tau /\hbar }$ , where $\widetilde{k}=l_{B}k$ . Integrating
Eq.(\ref{d/dtI_+appr}) over $t$ , and solving for $\widetilde{I}_{+}\left(
t\right) $ , the time dependence of the coherence $I_{+}$ is given by 
\[
I_{+}\left( t\right) =I_{+}\left( 0\right) J\left( t\right) 
\]
where $J\left( t\right) =e^{i\Omega \left( t\right) -\Gamma \left( t\right)
} $ , and 
\begin{equation}
\Gamma \left( t\right) =\left( \frac{C\left| u\left( 0\right) \right| ^{2}}{%
4\pi l_{B}^{2}l}\right) ^{2}\int_{0}^{\infty }\widetilde{k}d\widetilde{k}e^{-%
\frac{1}{2}\widetilde{k}^{2}}\frac{1-\cos \left[ \frac{1}{\hbar }%
E_{ex}\left( k\right) -\omega \right] t}{\left[ E_{ex}\left( k\right) -\hbar
\omega \right] ^{2}}  \label{Gamma}
\end{equation}

This result is similar to the expression found by Palma et al. \cite{Palma96}
in an artificial model of pure decoherence, i.e. when energy transfer
between the qbit and its environment is not allowed. The remarkable feature
of this expression is due to the presence of the energy gap $\varepsilon
_{sp}$ in the spin exciton spectrum, which is typically much larger than the
nuclear Zeeman energy $\hbar \omega $ . This guarantees that the denominator
in the integrand in Eq.(\ref{Gamma}) is always larger than or equal to $%
\varepsilon _{sp}^{2}$ , and that for times $t$ $\gg $ $\hbar /\varepsilon
_{sp}$ , 
\[
\Gamma \left( t\right) \longrightarrow \widetilde{A}^{2}\int_{0}^{\infty }%
\frac{\widetilde{k}d\widetilde{k}e^{-\frac{1}{2}\widetilde{k}^{2}}}{\left[ 
\widetilde{E}_{ex}\left( k\right) \right] ^{2}} 
\]
with the dimensionless exciton energy $\widetilde{E}_{ex}\left( k\right)
=E_{ex}\left( k\right) /\varepsilon _{sp}\geq 1$, and hyperfine coupling
constant $\widetilde{A}=\frac{C\left| u\left( 0\right) \right| ^{2}}{4\pi
l_{B}^{2}l\varepsilon _{sp}}$.

Thus , we find that during a short time scale , of the order of $\hbar
/\varepsilon _{sp}$ , the coherence , $I_{+}\left( t\right) $ , of a single
nuclear spin diminishes and then saturates for a very long time ( i.e. of
the order of the relaxation time $T_{2}$ , see below) at $I_{+}\left(
0\right) e^{-\varkappa \widetilde{A}^{2}}$, where $\varkappa
=\int_{0}^{\infty }\frac{\widetilde{k}d\widetilde{k}e^{-\frac{1}{2}%
\widetilde{k}^{2}}}{\left[ \widetilde{E}_{ex}\left( k\right) \right] ^{2}}%
\sim \frac{2\varepsilon _{sp}}{\varepsilon _{c}}$ . For $%
GaAs/Al_{x}Ga_{1-x}As$ heterostructure the coupling constant $\widetilde{A}$
is typically of the order of $10^{-4}$ \ \ [\cite{BMV95}]\ , implying an
extremely small deviation , i.e. $\sim \varkappa \widetilde{A}^{2}\sim
10^{-9}$, from a pure state of a single nuclear spin.

It is interesting to compare this effect to the decoherence caused to a
nuclear spin as a result of the scattering of spin excitons by random
impurities. This mechanism leads to a complete decoherence within a time
scale [\cite{MPV00}] 
\[
T_{2}\sim \frac{1}{\widetilde{A}^{2}U_{2}}\hbar /\varepsilon _{sp} 
\]
\ where $U_{2}\equiv \frac{1}{2\pi l_{B}^{2}\varepsilon _{c}^{2}}\int d^{2}r$
$\left\langle U_{imp}({\bf r})U_{imp}\left( {\bf r}^{\prime }\right)
\right\rangle $ is a dimensionless correlator of the impurity potential $%
U_{imp}({\bf r})$ [\cite{IMV91}]. In $GaAs/Al_{x}Ga_{1-x}As$
heterostructures $\hbar /\varepsilon _{sp}\sim 10^{-12}\sec $ and $U_{2}$ is
typically $\sim 0.001$[\cite{MPV00}],\ so that $T_{2}\sim 0.1\sec $.

Thus, due to the extreme weakness of the hyperfine contact interaction with
the 2D electron gas under high magnetic fields the decoherence of a single
qbit arising from both impurity scattering and quantum fluctuations in such
a system is extremely small. A coherent superposition of a great number of
qbits can therefore survive in the computer memory during a very long time
period $t\ll T_{2}$. To find an upper bound for the length of such a memory
let us consider $N$ independent nuclear spins located at various positions $%
{\bf r}_{j}$ in the quantum well. A number, $n$ , stored in the memory,
corresponds to the direct product of $N$ pure nuclear spin states $\left|
n\right\rangle =$ $\left| n_{1}\right\rangle $ $\otimes \left|
n_{2}\right\rangle \otimes ...\otimes \left| n_{N}\right\rangle $ , where $%
\left| n_{j}\right\rangle =\sum_{\sigma =\pm 1}\delta _{n_{j},\sigma }\left|
j,\sigma \right\rangle $ , $\delta _{n_{j},\sigma }$ is the Kronecker delta
, and $\left| j,\sigma \right\rangle $ is a nuclear state with spin
projection $\sigma $ for a nucleus located at ${\bf r}_{j}$.

To carry out a quantum computing process, however, a coherent superposition
of such products , \ i.e. $\left| \psi \right\rangle =\sum_{n=1}^{N}\alpha
_{n}\left| n\right\rangle $ \ (see e.g. \cite{Unruh95}), should be prepared
at time $t=0$. \ \ This superposition may be represented more transparently
for our purposes by the direct product of $N$ mixed spin up and spin down
states, 
\begin{equation}
\left| \psi \left( t=0\right) \right\rangle =\prod_{j=1}^{N}\otimes \left(
u_{j}\left| j,\uparrow \right\rangle +v_{j}\left| j,\downarrow \right\rangle
\right)  \label{Psi}
\end{equation}
with the normalization $\left| u_{j}\right| ^{2}+\left| v_{j}\right| ^{2}=1$
. \ 

Let us assume that at time $t=-t_{0}<0$ the filling factor was tuned to a
fixed value $\nu =\nu _{0}\neq 1$ and then kept constant until $t=0$ . If $%
t_{0}\gg T_{2}\left( \nu _{0}\right) $ then at $t=0$ the nuclear spin system
is in the ground state corresponding to the 2D electron system at $\nu =\nu
_{0}$. Suppose that at time $t=0$ the filling factor is quickly switched (
i.e. on a time scale much shorter than $T_{2}\left( \nu _{0}\right) $ ) back
to $\nu =1$ so that the nuclear spin system is suddenly trapped in its
instantaneous configuration corresponding to $\nu =\nu _{0}\neq 1$ . Thus
the nuclear spins will find themselves for a long time $t$ ( $\gg $ $%
T_{2}\left( \nu _{0}\right) $ ) almost frozen in the ground state
corresponding to the 2D electron system at $\nu =\nu _{0}$, since $%
T_{2}\left( \nu =1\right) \gg T_{2}\left( \nu _{0}\right) $.

The corresponding state of the nuclear spin system can be found by
considering $u_{j}$ and $v_{j}$ in Eq.(\ref{Psi}) as variational parameters
, and then minimizing the energy functional ${\cal E=}\left\langle \psi
\right| {\cal H}\left| \psi \right\rangle $ ,

\[
{\cal H}=-\gamma _{n}\sum_{j=1}^{N}\widehat{{\bf I}}_{j}\cdot {\bf B}%
_{0}+A\sum_{j=1}^{N}{\bf S}\left( {\bf r}_{j}\right) \cdot \widehat{{\bf I}}%
_{j}
\]
with respect to $u_{j}$ , $v_{j}$ . \ Note that the effective nuclear spin
Hamiltonian ${\cal H}$ is obtained after averaging over the electronic
('bath') states, so that ${\bf S}\left( {\bf r}_{j}\right) =\left\langle 
\widehat{{\bf S}}\left( {\bf r}_{j}\right) \right\rangle \approx
\left\langle 0\right| \widehat{{\bf S}}\left( {\bf r}_{j}\right) \left|
0\right\rangle $ is the expectation value of the electronic spin density at
the nuclear position ${\bf r}_{j}$. \ As noted above , at $\nu _{0}\neq 1$ ,
this density has nonzero transverse components, associated with the
skyrmionic spin texture, smoothly varying in space.

A simple calculation shows that ${\cal E=}\sum_{j}\varepsilon \left(
u_{j},v_{j}\right) $ , 
\[
\varepsilon \left( u_{j},v_{j}\right) =\frac{1}{2}\omega _{j}\left( \left|
v_{j}\right| ^{2}-\left| u_{j}\right| ^{2}\right) +\frac{1}{2}\left[
Av_{j}u_{j}^{\ast }S_{-}\left( {\bf r}_{j}\right) +c.c\right] 
\]
where $\omega _{j}=\gamma _{n}B_{0}-AS_{z}\left( {\bf r}_{j}\right) $ is the
local nuclear Zeeman energy. The extremum conditions ( subject to the
normalization $\left| u_{j}\right| ^{2}+\left| v_{j}\right| ^{2}=1$ ) $\frac{%
\partial \varepsilon }{\partial u_{j}^{\ast }}-\epsilon _{j}u_{j}=0$ , $%
\frac{\partial \varepsilon }{\partial v_{j}^{\ast }}-\epsilon _{j}v_{j}=0$
are readily solved to yield: 
\begin{equation}
\left| u_{j}\right| ^{2}=\frac{1}{2}\left( 1\pm \frac{\omega _{j}}{\sqrt{%
2\epsilon _{j}}}\right) \;,\;\;\left| v_{j}\right| ^{2}=\frac{1}{2}\left(
1\mp \frac{\omega _{j}}{\sqrt{2\epsilon _{j}}}\right)   \label{uvequa}
\end{equation}
where 
\[
\epsilon _{j}=\frac{1}{2}\sqrt{\omega _{j}^{2}+A^{2}S_{-}\left( {\bf r}%
_{j}\right) S_{+}\left( {\bf r}_{j}\right) }
\]

In this state the nuclear spin polarization $\left\langle \psi \right| 
\widehat{{\bf I}}_{j}\left| \psi \right\rangle $ follows the underlying
electronic spin texture; the transverse component takes the form 
\[
\left\langle \psi \right| \widehat{I}_{j,+}\left| \psi \right\rangle
=u_{j}^{\ast }v_{j}=AS_{+}\left( {\bf r}_{j}\right) /2\epsilon _{j} 
\]
whereas the longitudinal component is 
\[
\left\langle \psi \right| \widehat{I}_{j,z}\left| \psi \right\rangle =\frac{1%
}{2}\left( \left| u_{j}\right| ^{2}-\left| v_{j}\right| ^{2}\right) =\pm
\omega _{j}/2\epsilon _{j} 
\]

The topological rigidity of the skyrmionic spin texture thus ensures the
rigidity of the coherences $u_{j}^{\ast }v_{j}$ over a large spatial region.

The key parameter here is the mixing parameter $\eta _{j}\equiv A^{2}\left|
S_{+}\left( {\bf r}_{j}\right) \right| ^{2}/\omega _{j}^{2}$ , which becomes
significant when the transverse component $S_{+}\left( {\bf r}_{j}\right) $
does not vanish over a large spatial region , as is the case for skyrmion
spin texture. For very small mixing parameter $\eta _{j}$ one finds a pure
nuclear ferromagnetic state, namely $\left| u_{j}\right| ^{2}=1$ , $\left|
v_{j}\right| ^{2}=0$ , corresponding to the ground state , or $\left|
u_{j}\right| ^{2}=0$ , $\left| v_{j}\right| ^{2}=1$ , corresponding to the
fully saturated nuclear spin system. In the opposite limit of very strong
mixing ( $\eta _{j}\gg 1$ ) $\left| u_{j}\right| ^{2}=\left| v_{j}\right|
^{2}=\frac{1}{2}$ , which is the desired state for quantum computing \cite
{Unruh95}, all the numbers $n$ are stored in the memory with equal
probability. \ It should be noted that, since $\left| S_{+}\right| \leq 1$ ,
large values of $\eta _{j}$ can be obtained only when the nuclear Zeeman
energy $\omega _{j}$ is much smaller than the hyperfine coupling constant $A$
.

Now, during the long time following $t=0$ , when the electronic system is
set at filling factor $\nu =1$ so that its ground state is a uniform quantum
ferromagnet and the elementary spin excitations are the spin waves discussed
above , the nuclear state $\left| \psi \left( t\right) \right\rangle $
evolving from $\left| \psi \left( 0\right) \right\rangle $ after time $t$
can be readily calculated in terms of the operators $\widehat{I}_{j,\pm
}\left( t\right) ,\widehat{I}_{j,z}\left( t\right) $ . Using the solutions $%
\widehat{I}_{j,+}\left( t\right) =J\left( t\right) \widehat{I}_{j,+}\left(
0\right) $,$\widehat{I}_{j,z}\left( t\right) =I\left( t\right) \widehat{I}%
_{j,z}\left( 0\right) $ , derived above ( Eq.(\ref{Gamma}) ), it is easy to
show that the probability that after time $t$ the memory remains in the
coherent state $\psi \left( 0\right) $: 
\begin{eqnarray}
P_{\psi } &=&\left| \left\langle \psi \left( 0\right) \mid \psi \left(
t\right) \right\rangle \right| ^{2}  \label{Prob} \\
&=&\frac{1}{2^{N}}\prod_{j=1}^{N}\{\left[ 1+I\left( t\right) \right] +4\left[
I\left( t\right) -%
\mathop{\rm Re}%
J\left( t\right) \right] \left[ \left| u_{j}\right| ^{4}-\left| u_{j}\right|
^{2}\right] \}  \nonumber
\end{eqnarray}

\bigskip Let us further assume that the mixing parameter $\eta _{j}$ are
large so that in the initial state $\psi \left( 0\right) $ , $\ \left|
u_{j}\right| ^{2}=1/2$ as required. Under these circumstances we easily find
that:

\[
P_{\psi }\left( t\right) =\left[ \frac{1+%
\mathop{\rm Re}%
J\left( t\right) }{2}\right] ^{N}\approx \exp \left\{ -\frac{1}{2}N\left[ 1-%
\mathop{\rm Re}%
J\left( t\right) \right] \right\} \approx e^{-\frac{1}{2}N\Gamma \left(
t\right) } 
\]

It is evident that due to the even distribution of nuclear spins in the
initial state $\psi \left( 0\right) $ the survival probability $P_{\psi
}\left( t\right) $ depends only on the decoherence factor $J\left( t\right) $%
. The decay of $P_{\psi }\left( t\right) $ therefore follows $e^{-\frac{1}{2}%
N\Gamma \left( t\right) }$ , saturating at $e^{-\frac{1}{2}N\varkappa 
\widetilde{A}^{2}}$ for $t$ $\gg $ $\hbar /\varepsilon _{sp}$ . Note that,
despite the much larger drop in the level of coherence in the many qbit
system , \ the time scale over which the coherence diminishes is the same as
for a single qbit.

An upper bound on the length of possible memories in future quantum
computers based on the proposed model can be now estimated by the
requirement $P_{\psi }\left( t\gg \hbar /\varepsilon _{sp}\right) \sim 1/e$
, which yields 
\[
N_{\max }\sim 2/\widetilde{A}^{2} 
\]
The restriction imposed by this condition is inherent to the manipulation
mechanism of qbits via the hyperfine interaction with the electron spins,
and so can not be removed or even relaxed by any technical improvement.

\bigskip In conclusion we have found that a system of many nuclear spins,
coupled to the electronic spins in the 2D electron gas through the Fermi
contact hyperfine interaction, partially looses its phase coherence during
the short (electronic) time $\hbar /\varepsilon _{sp}$, \ even under the
ideal conditions of the QHE, where both $T_{1}$, and $T_{2}$ are infinitely
long. The effect arises as a result of vacuum quantum fluctuations
associated with virtual excitations of spin waves (or spin excitons ) by the
nuclear spins. The incompleteness of the resulting decoherence is due to the
large energy gap (on the scale of the nuclear Zeeman energy) of these
excitations whereas the extreme weakness of the hyperfine interaction with
the 2D electron gas under high magnetic fields guarantees that the loss of
coherence of a single nuclear spin is extremely small. The memory of a
quantum computer to be constructed in such a system is therefore limited in
principle to lengths of the order of $N_{\max }$ , which is found to be
about $10^{9}$ for GaAS multiple quantum well structure.

\bigskip

This research was supported by the German-Israeli Foundation for Scientific
Research and Development, Grant No. G-0456-220.07/95, and by the fund for
the promotion of research at the Technion. Yu.A.B. wishes to acknowledge
support from grants RFFI-00-02-17292, IR-97-0076, and INTAS-99-01146. \ I.V.
acknowledges the support by the Caesarea Edmond Benjamin de Rotschild
Foundation.

\end{document}